\documentclass[12pt,fleqn]{article}
\usepackage{graphicx}

\usepackage{latexsym}

\usepackage{amsmath}
\usepackage{amsthm}
\usepackage{amssymb}
\usepackage{amsfonts}

\numberwithin{equation}{section}

\begin{document}

\newcommand{\rf}[1]{(\ref{#1})}
\newcommand{\rff}[2]{(\ref{#1}\ref{#2})}

\newcommand{\ba}{\begin{array}}
\newcommand{\ea}{\end{array}}

\newcommand{\be}{\begin{equation}}
\newcommand{\ee}{\end{equation}}

\newcommand{\const}{{\rm const}}
\newcommand{\ep}{\varepsilon}
\newcommand{\Cl}{{\cal C}}
\newcommand{\rr}{{\vec r}}
\newcommand{\ph}{\varphi}
\newcommand{\R}{{\mathbb R}}  
\newcommand{\C}{{\mathbb C}}  
\newcommand{\T}{{\mathbb T}}  
\newcommand{\Z}{{\mathbb Z}}  

\newcommand{\e}{{\bf e}}

\newcommand{\m}{\left( \ba{c}}
\newcommand{\ema}{\ea \right)}
\newcommand{\mm}{\left( \ba{cc}}
\newcommand{\miv}{\left( \ba{cccc}}

\newcommand{\scal}[2]{\mbox{$\langle #1 \! \mid #2 \rangle $}}
\newcommand{\ods}{\par \vspace{0.5cm} \par}
\newcommand{\dis}{\displaystyle }
\newcommand{\mc}{\multicolumn}

\newtheorem{prop}{Proposition}
\newtheorem{Th}[prop]{Theorem}
\newtheorem{lem}{Lemma}
\newtheorem{rem}{Remark}
\newtheorem{cor}[prop]{Corollary}
\newtheorem{Def}{Definition}
\newtheorem{open}{Open problem}
\newtheorem{ex}{Example}
\newtheorem{exer}{Exercise}

\newenvironment{Proof}{\par \vspace{2ex} \par
\noindent \small {\it Proof:}}{\hfill $\Box$ 
\vspace{2ex} \par }

\title{\bf 
On the exact discretization of the classical harmonic oscillator equation}
\author{
 {\bf Jan L.\ Cie\'sli\'nski}\thanks{\footnotesize
 e-mail: \tt janek\,@\,alpha.uwb.edu.pl}
\\ {\footnotesize Uniwersytet w Bia{\l}ymstoku,
Wydzia{\l} Fizyki}
\\ {\footnotesize ul.\ Lipowa 41, 15-424
Bia{\l}ystok, Poland}
}

\date{}

\maketitle

\begin{abstract}
We discuss the exact discretization of the classical harmonic oscillator equation (including the inhomogeneous case and multidimensional generalizations) with a special stress on the energy integral. We present and suggest some  numerical applications. 
\end{abstract}

\ods



\pagebreak

\section{Introduction}

By the exact discretization of an ordinary differential equation $\dot x = f (x)$, where $x (t) \in \R^N$, we mean the difference equation $X_{n+1} = F (X_n)$, where $X_n \in \R^N$, such that $X_n = x (t_n)$. Initial data coincide as well, because $X_0 = x (t_0)$. 
Exact discretizations have been first studied by Potts \cite{Po} and a  detailed account of subsequent developments can be found in Agarwal's book \cite{Ag}. 
It is worthwhile to point out that all linear ordinary differential equations with constant coefficients admit exact discretizations \cite{Ag,Po}. 

In this paper we discuss exact discretizations for the harmonic oscillator and its various extensions, including multidimensional, damped and inhomogeneous cases. In other words, we consider the exact discretization of  motions in a quadratic potential.

Discretizations are closely related to numerical finite differnce schemes. Some numerical algoritms are known to integrate exactly the harmonic oscillator equation, e.g., the so-called Gautschi-type methods \cite{Gau,HoL2,SB} and exponential integrators \cite{Law,MW,Pope}. 
Recently, the problem of the exact discretization of the harmonic oscillator equation with the constant force turns out to be very important for the construction of new numerical algorithms (``localy exact numerical schemes'' \cite{Ci-focm,CR-long,CR-grad}, see also Section~\ref{sec-loc-exact}) and this is our main motivation for studying the subject of exact discretizations.

\section{Exact discretization of the one-dimensional harmonic oscillator}

It is well known \cite{Ag,CR-ade,Po} that the discrete second-order linear equation
\be  \label{ex-osc}
\frac{x_{n+1} - 2 x_n + x_{n-1}}{\left( \frac{2}{\omega} \sin \frac{\ep \omega}{2} \right)^2 } 
+ \omega^2 x_n  = 0 
\ee
discretizes exactly the classical harmonic oscillator equation 
\be  \label{osc}
 \ddot x + \omega^2 x = 0 \ , 
\ee
where $x = x (t)$ and the dot denotes the derivative with respect to $t$. 
In other words, any solution $x_n$ of \rf{ex-osc} can be expressed as $x_n = x (\ep n)$, where  the function $t \rightarrow x (t)$ satisfies the equation \rf{osc} (the time step $t_{n+1} - t_n = \ep$ is constant). Defining $v_n := \dot x (\ep n)$ we derive 
the exact discretization of the velocity $\dot x (t)$ 
\be  \label{vn} 
  v_n = \frac{x_{n+1} - x_n \cos\omega\ep}{\frac{\sin\omega\ep}{\omega}} \ , 
\ee
compare \cite{CR-ade} (see also below, where more general cases  will be presented in detail).

\subsection{Equivalent forms of the exact discretization} 

The exact discretization of the harmonic oscillator equation can be represented in several equivalent forms which are useful in various contexts. 
The simplest form, following immediately from \rf{ex-osc}, is
\be  \label{osc-cos}
x_{n+1} - 2 \cos\omega\ep \ x_n + x_{n-1} = 0 \ .
\ee
Equations similar to \rf{osc-cos} can be found in \cite{Hild} (Section 1.7, Examples 2 and 5) without mentioning their relations with exact discretizations.

\begin{prop} 
The equation \rf{ex-osc} can be represented in the following equivalent form:
\be  \label{osc-dd}
\frac{ \dis \left( \frac{x_{n+1} - x_n \cos\omega\ep}{\frac{\sin\omega\ep}{\omega}} \right)  - \left( \frac{x_{n} - x_{n-1} \cos\omega\ep}{\frac{\sin\omega\ep}{\omega}} \right) \cos\omega\ep }{\dis \frac{\sin\omega\ep}{\omega} } + \omega^2 x_{n-1} = 0 \ .
\ee
\end{prop}
\begin{Proof} 
A simple straightforward calculation.
\end{Proof}

\begin{cor}
Defining a new difference operator 
\be  \label{newDelta} 
  \Delta_\ep := \frac{T - \cos\omega\ep}{\omega^{-1} \sin\omega\ep} \ , 
\ee
where $T$ is the shift operator (i.e., $ (T x)_n = x_{n+1}$), we can rewrite \rf{osc-dd} as 
\be
\Delta_\ep^2 x + \omega^2 x = 0 \ . 
\ee
\end{cor}

Note that in the limit $\ep\rightarrow 0$ the equation \rf{ex-osc} assumes the form of the symmetric Euler finite difference scheme 
\be  \label{Eu-osc}
\frac{x_{n+1} - 2 x_n + x_{n-1}}{\ep^2 } 
+ \omega^2 x_n  = 0 \ , 
\ee
while the equation \rf{osc-dd} apparently tends to the forward Euler finite difference scheme:  
\be  \label{osc-dd1}
\frac{ \dis \left( \frac{x_{n+1} - x_n }{\ep} \right)  - \left( \frac{x_{n} - x_{n-1} }{\ep} \right) }{\ep} + \omega^2 x_{n-1} = 0 \ 
\ee
(when $\cos\omega\ep$ is approximated by $1$). However, performing this limit with full care (i.e., leaving the second term $\frac{1}{2} \omega^2 \ep^2$ in the Taylor expansion of $\cos\omega\ep$) we also get \rf{Eu-osc}.  

The harmonic oscillator equation \rf{osc} can be represented in the Hamiltonian form
\be \label{osc-Ham}
m \dot x = p, \quad \dot p = - k x \ , 
\ee
where $\omega^2 = \frac{k}{m}$. 

\ods
\begin{prop}  \label{prop-dis-Ham}
The exact discretization of \rf{osc-Ham} is given by
\be
\ba{l}  \dis  \label{osc-Ham-dis}
\frac{ m (x_{n+1} - x_n ) }{\delta } = \frac{p_{n+1} + p_n}{2} \ ,  \\[3ex] \dis
\frac{ p_{n+1} - p_n }{\delta} = - \frac{k ( x_{n+1} + x_n )}{2} \ , 
\ea \ee
where 
\be  \label{deltaep} 
\delta = \frac{2}{\omega} \tan \frac{\omega \ep}{2} \ .
\ee
\end{prop}

\begin{Proof} It is enough to show that the system \rf{osc-Ham-dis}, where $p_n = m v_n$, is equivalent to the system \rf{vn}, \rf{osc-cos}. 
Solving \rf{osc-Ham-dis} with respect to $p_n$ and $p_{n+1}$, and then shifting index $n+1 \rightarrow n$ in the first resulting equation, we get
\be \ba{l} \dis   \label{pny}
p_{n+1} = \frac{m}{\delta} (x_{n+1} - x_n) - \frac{k \delta}{4} (x _{n+1} + x_n ) \ , \\[3ex] \dis
p_n = \frac{m}{\delta} (x_{n+1} - x_n) + \frac{k \delta}{4} (x_{n+1} + x_n ) \ , \\[3ex] \dis
 p_n = \frac{m}{\delta} (x_n - x_{n-1}) - \frac{k \delta}{4} (x_n + x_{n-1}) \ .
\ea \ee
Eliminating $p_n$ from the last two equations we get
\be
\left( \frac{m}{\delta} + \frac{k \delta}{4} \right) x_{n+1} - 2 
\left( \frac{m}{\delta} - \frac{k \delta}{4} \right) x_n + \left( \frac{m}{\delta} + \frac{k \delta}{4} \right) x_{n-1} = 0 \ , 
\ee
which is equivalent to \rf{osc-cos} if and only if 
\be
\cos\omega\ep = \frac{4m - k \delta^2}{4 m + k \delta^2} \ ,  \quad  {\rm i.e.,} \quad \frac{k \delta^2}{4 m} = \tan^2 \frac{\omega \ep}{2} \ , 
\ee
which is satisfied by virtue of \rf{deltaep}. Substituting \rf{deltaep} and $k = m \omega^2$ into the middle equation of \rf{pny} we get
\be  \label{pncos}
 p_n = \frac{ m \omega ( x_{n+1}  - x_n \cos\omega\ep )}{\sin\omega\ep} \ ,
\ee
i.e., $p_n = m v_n$, where $v_n$ is given by \rf{vn}, which ends the proof. 
\end{Proof}

Taking into account \rf{pncos} and the first equation of \rf{pny} we can solve the system \rf{osc-Ham-dis} with respect to $x_{n+1}, p_{n+1}$ obtaining 
\be  \label{dis-osc-explicit}
\m x_{n+1} \\ p_{n+1} \ema = \mm \cos\omega\ep &  (m \omega)^{-1} \sin\omega\ep \\  - m \omega \sin\omega\ep & \cos\omega\ep \ema \m x_n \\ p_n \ema \ .
\ee

\subsection{Discrete analogues of the energy integral}

The harmonic oscillator equation $\ddot x + \omega^2 x = 0$ 
has the following integral of motion 
\be
  \frac{1}{2} {\dot x}^2 + \frac{1}{2} \omega^2 x^2 =  E = \const \ , 
\ee
where $E$ can be interpreted as the energy per a mass unit. 
Trying to find a conservation law for its discrete analogue, \rf{ex-osc}, one may consider the following quantities: 
\be  \ba{l} \label{Ekn} \dis
E^{(0)}_n = x_{n+1}^2 - 2 \cos\omega\ep \ x_n x_{n+1} + x_n^2  \ , \\[4ex] \dis 
E^{(1)}_n := \frac{1}{2} \left( \frac{x_{n+1} -  x_n}{ \frac{2}{\omega} \sin \frac{\ep \omega}{2} } \right)^2  
+ \frac{1}{2} \omega^2 x_n x_{n+1}  \ , \\[4ex]\dis
E^{(2)}_n := \frac{1}{2} \left( \frac{x_{n+1} - x_n \cos\omega\ep}{\frac{\sin\omega\ep}{\omega}} \right)^2  + \frac{1}{2} \omega^2 x_n^2 \ \equiv \  \frac{1}{2} (\Delta_\ep x_n)^2 + \frac{1}{2} \omega^2 x_n^2 \ .
\ea \ee

\begin{prop} 
We assume that $x_n$ satisfies \rf{ex-osc}. Then, 
$E^{(k)}_{n+1} = E^{(k)}_n$ for any $k = 0,1,2$ and any $n \in \Z$. Moreover,  
\be  \label{identE} \ba{l}  \dis
E^{(1)}_n = \frac{1}{2} \left( \frac{\omega}{2 \sin\frac{\omega\ep}{2}} \right)^2 \ E^{(0)}_n \ , \\[4ex]  \dis
E^{(2)}_n = \left( 1 + \tan^2 \frac{\omega\ep}{2} \right) E^{(1)}_n = \frac{1}{2} \left( \frac{\omega}{ \sin{\omega\ep}} \right)^2 \ E^{(0)}_n \ .
\ea \ee
\end{prop}

\begin{Proof}
The identities \rf{identE} can be easily checked in a straightforward way. Therefore it is enough to show that $E^{(0)}_{n} = E^{(0)}_{n-1}$. We compute: 
\[
E^{(0)}_{n} - E^{(0)}_{n-1} = (x_{n+1} - x_{n-1})(x_{n+1} + x_{n-1} - 2 \cos\omega\ep \ x_n ) \ , 
\]
which vanishes provided that \rf{osc-cos} holds. 
\end{Proof}

Therefore all three quantities defined by formulas \rf{Ekn} are integrals of motion, or, more precisely, they are different representations of the discrete analogue of the energy integral.  
In a sense these three forms of the energy integral correspond to the above three  equivalent forms of the exact discrete harmonic oscillator equation. 
Namely, $E^{(0)}$ can be obtained as a quadratic integral of motion for \rf{osc-cos}. Then, $E^{(1)}$ is a natural guess in the spirit of Mickens' approach \cite{Mic}, strongly implied by the form \rf{ex-osc}. Finally, $E^{(2)}$ is the best  discrete analogue of the energy integral. Indeed, 
having the exact discretization we expect that there exists a discrete analogue of this conservation law, i.e., 
\be  \label{En2}
   E_n = \frac{1}{2} v_n^2 + \frac{1}{2} \omega^2 x_n^2 
\ee
where $v_n$ has to be defined. Note that the invariant $E^{(2)}$ is of the form \rf{En2} provided that $v_n$ is defined by \rf{vn}. 

\subsection{A family of geometric integrators}

In this subsection we consider a family of discrete systems  
containing the exact discretization of the harmonic oscillator.  

\begin{prop}  \label{prop-sympl} 
The discrete map $(x_n, p_n) \rightarrow (x_{n+1}, p_{n+1})$, defined by: 
\be  \label{map}
 x_{n+1} - \gamma x_n + x_{n-1} = 0 \ , \quad p_n = \alpha x_{n+1} - \beta x_n \ , 
\ee
is both symplectic and energy-preserving for any choice of $\alpha, \beta, \gamma$. If, moreover, the coefficients depend on the time step $\ep$, satisfying constraints  
\be \label{abg}
\alpha (-\ep) = - \alpha (\ep) \ , \qquad \gamma (\ep) = \frac{\beta (\ep) - \beta (-\ep) }{\alpha (\ep)} \ , 
\ee
then the map is also symmetric (time-reversible).
\end{prop}

\begin{Proof} We compute \ 
$p_{n+1} = \alpha (\gamma x_{n+1} - x_n) - \beta x_{n+1}$. Hence, 
\[  \ba{l}
  d x_{n+1} \wedge d p_{n+1} =  - \alpha d x_{n+1} \wedge d x_n  \ , 
\\[1ex]
d x_n \wedge d p_n = \alpha d x_n \wedge d x_{n+1} \ . 
\ea \]
Thus $d x_{n+1} \wedge d p_{n+1} = d x_n \wedge d p_n $ which means that the map is symplectic (and volume-preserving) \cite{HLW}. An integral of the discrete evolution can be easily found by multiplying the first equation of \rf{map} by $x_{n+1} - x_{n-1}$. Indeed, 
\[
0 = ( x_{n+1} - \gamma x_n + x_{n-1} )(x_{n+1} - x_{n-1}) \equiv 
x_{n+1}^2 - \gamma x_{n+1} x_{n} + \gamma x_n x_{n-1} - x_{n-1}^2 \ ,  
\]
which means that
\be
x_{n+1}^2 - \gamma x_{n+1} x_n + x_n^2 = x_n^2 - \gamma x_n x_{n-1} + x_{n-1}^2 = \const \ .
\ee
In order to find conditions for time-reversibility \cite{HLW}, we rewrite \rf{map} as follows:
\renewcommand{\arraystretch}{1.5}
\[ 
\m x_{n+1} \\  p_{n+1} \ema = A (\ep) \m x_n \\ p_n \ema \ , \qquad A (\ep) = \mm \frac{\beta}{\alpha} & \frac{1}{\alpha} \\ \gamma \beta - \frac{\alpha^2 + \beta^2}{\alpha}  & \gamma - \frac{\beta}{\alpha}  \ema \ , 
\]
where $\alpha  = \alpha (\ep)$, $\beta = \beta (\ep)$ and $\gamma = \gamma (\ep)$. 
Computing $A^{-1}$ we get: 
\[ 
\m x_n \\ p_n \ema  = A^{-1} (\ep) \m x_{n+1} \\  p_{n+1} \ema \ , \qquad A^{-1} (\ep) = \mm \gamma - \frac{\beta}{\alpha}  & - \frac{1}{\alpha} \\ \frac{\alpha^2 + \beta^2}{\alpha} - \gamma \beta   & \frac{\beta}{\alpha} \ema  \ .
\]
\renewcommand{\arraystretch}{1}
Time-reversibility means that $A (-\ep) = A^{-1} (\ep)$, i.e., 
\be  \ba{l}
\alpha (-\ep) = - \alpha (\ep) \ , \\[3ex]
\gamma (\ep) - \frac{\beta (\ep)}{\alpha (\ep)} = \frac{\beta ( -\ep)}{\alpha (-\ep)} \ , \\[3ex]
\gamma (\ep) \beta (\ep) - \frac{\alpha^2 (\ep) + \beta^2 (\ep)}{\alpha (\ep)} = \frac{\alpha^2 (- \ep) + \beta^2 (-\ep)}{\alpha (-\ep)} - \gamma (-\ep) \beta (-\ep) \ . 
\ea \ee
The first two equations yield \rf{abg}. The third equation is identically satisfied provided that the first two equations hold.  
\end{Proof}

The maps defined by \rf{map} contain a large family of geometric numerical integrators for the classical harmonic equation \rf{osc}. Indeed, if 
\be
\lim_{\ep\rightarrow 0} \ep \alpha (\ep) = m \ , \quad 
\lim_{\ep\rightarrow 0} \ep \beta (\ep) = m \ , \quad 
\lim_{\ep\rightarrow 0} \frac{2 - \gamma (\ep)}{\ep^2} = \frac{k}{m} \equiv \omega^2 \ , 
\ee
then the continuum limit of \rf{map} (divided by $\ep^2$) is given by $\ddot x + \omega^2 x = 0$ and $p = m \dot x$.  
In particular, the exact discretization of the harmonic oscillator is characterized by 
\be
\alpha (\ep) = \frac{m\omega}{\sin\omega\ep} \ , \quad \beta (\ep) = m \omega \cot (\omega \ep) \ , \quad \gamma (\ep) = 2 \cos\omega\ep \ . 
\ee

\begin{cor}
The exact discrtization of the harmonic oscillator equation satisfies all assumptions of Proposition~\ref{prop-sympl} and, therefore, is symplectic, volume-preserving, energy-preserving and time-reversible. 
\end{cor}

\section{Harmonic oscillator with a constant driving force}

In this section we consider the exact discretization 
of the classical harmonic oscillator equation with a constant force:
\be  \label{osc-force} 
\ddot x + \omega^2 x = g \ , 
\ee
where $\omega$ and $g$ are constant. This problem deserves special attention in the context of some numerical applications, see Section~\ref{sec-numer}. 

\subsection{Variable time step}

The general exact solution of the equation \rf{osc-force} and its derivative  read 
\be \ba{l} \dis
x (t) = A \cos \omega t + B \sin \omega t + \frac{g}{\omega^2} \ , \\[2ex] \dis
\dot x (t) = \omega B \cos \omega t - \omega A \sin\omega t \ .
\ea \ee
We consider the exact discretization
\be \ba{l} \dis  \label{exact11}
x_n := A \cos(\omega t_n )  + 
B \sin(\omega t_n ) + \frac{g}{\omega^2} \ , \\[2ex] \dis
v_n := \omega B \cos \omega t_n - \omega A \sin \omega t_n \ .
\ea \ee

\ods

\begin{prop}
The exact discretization (with a variable time step) of the equation \rf{osc-force} is given by
\be  \label{ab-force}
 a_{n-1} x_{n+1} - b_n x_n + a_n x_{n-1} = \frac{g}{\omega^2} \left( a_n + a_{n-1} - b_n \right) \ , 
\ee
where $a_n = \sin\omega\ep_n$, $b_n = \sin \omega (\ep_n + \ep_{n-1})$ and $\ep_n$ is a prescribed (variable) time step. 
\end{prop}

\begin{Proof}
Assuming the variable time step, $t_{n+1} - t_n = \ep_n$, and using \rf{exact11}, we have
\be \ba{l} \dis
x_{n+1} = A \cos(\omega t_n + \omega \ep_n)  + 
B \sin(\omega t_n + \omega \ep_n) + \frac{g}{\omega^2} \ , \\[3ex] \dis
x_{n-1} = A \cos(\omega t_n - \omega \ep_{n-1})  + 
B \sin(\omega t_n - \omega \ep_{n-1}) + \frac{g}{\omega^2} \ .
\ea \ee
Therefore
\be  \ba{l} \dis  \label{vn00}
x_{n+1} =  (x_n - \frac{g}{\omega^2} ) \cos\omega\ep_n + \sin\omega\ep_n ( B \cos\omega t_n - A \sin\omega t_n ) + \frac{g}{\omega^2}   , \\[3ex] \dis
x_{n-1} =  (x_n - \frac{g}{\omega^2} ) \cos\omega\ep_{n-1} - \sin\omega\ep_{n-1} ( B \cos\omega t_n - A \sin\omega t_n ) + \frac{g}{\omega^2}  , 
\ea \ee
and adding the first equation multiplied by $\sin\omega\ep_{n-1}$ to the second equation multiplied by $\sin\omega\ep_n$ we get
\[
\left( x_{n+1} - \frac{g}{\omega^2} \right) \sin\omega\ep_{n-1} + \left( x_{n-1} - \frac{g}{\omega^2} \right) \sin\omega\ep_n  = (x_n - \frac{g}{\omega^2} ) \sin (\omega \ep_n + \omega \ep_{n-1})  \ ,  
\]
which is equivalent to \rf{ab-force}.  
\end{Proof}

Comparing \rf{exact11} and the first equation of \rf{vn00}, we obtain 
\be  \label{vn-var-drv} 
  v_n = \frac{x_{n+1} - x_n \cos\omega\ep_n - \frac{g}{\omega^2} \left( 1 - \cos\omega\ep_n \right) }{ \frac{\sin\omega\ep_n}{\omega} } \ , 
\ee
which is the exact discretization of the velocity $\dot x$. 

\begin{prop}
The exact discretization of the harmonic oscillator with a constant force \rf{osc-force} is given by
\be  \label{dis-osc-explicit-g}
\m x_{n+1} \\ v_{n+1} \ema = \mm \cos\omega\ep_n  &  \omega^{-1}\sin\omega\ep_n \\  - \omega \sin\omega\ep_n  & \cos\omega\ep_n  \ema \m x_n \\ v_n \ema + \frac{g}{\omega^2} \m  1 - \cos\omega\ep_n   \\ 
\omega \sin\omega\ep_n  \ema  , 
\ee
where $v_n$ is the exact discretization of the velocity $\dot x$ and the time step $\ep_n$ is variable. 
\end{prop}

\begin{Proof}
From \rf{exact11} we derive:
\[ \ba{l}
x_{n+1} = (A \cos\omega t_n + B \sin\omega t_n) \cos\omega\ep_n 
+ ( B \cos\omega t_n - A \sin\omega t_n ) \sin\omega\ep_n + \omega^{-2} g  , \\[2ex]
v_{n+1} = - (A \omega \cos\omega t_n + B \omega \sin\omega t_n) \sin\omega\ep_n + ( B \omega \cos\omega t_n - A \omega \sin\omega t_n ) \cos\omega\ep_n   ,
\ea \]
Hence
\be\ba{l} 
x_{n+1} - \omega^{-2} g = (x_n - \omega^{-2} g) \cos\omega\ep_n + v_n \omega^{-1} \sin\omega\ep_n \ , \\[2ex]
v_{n+1} = - (x_n - \omega^{-2} g) \omega \sin\omega\ep_n + v_n \cos\omega\ep_n \ ,
\ea\ee
which is equivalent to \rf{dis-osc-explicit-g}.  
\end{Proof}

\begin{prop}
The exact discretization of the equation \rf{osc-force} can be represented in the following form, equivalent to \rf{dis-osc-explicit-g}:
\be \ba{l} \dis  \label{osc-graddel}
\frac{x_{n+1} - x_n}{\delta_n} = \frac{1}{2} \left( v_{n+1} + v_n \right) \ , \\[3ex]\dis
 \frac{v_{n+1} - v_n}{\delta_n} = - \frac{1}{2} \omega^2 \left( x_{n+1} + x_n \right) + g \ , 
\ea \ee
where 
\be  \label{deltan} 
\delta_n = \frac{2}{\omega} \tan \frac{\omega \ep_n}{2} \ .
\ee
\end{prop}

\begin{Proof} Analogical to the proof of Proposition~\ref{prop-dis-Ham}. 
\end{Proof}

\subsection{Constant time-step}

The equation \rf{ab-force} simplifies in the case $\ep_n = \ep = \const$. It is convenient to use the identity  
\be  
a_n + a_{n-1} - b_n \equiv 4 \sin \frac{\omega (\ep_n + \ep_{n-1})}{2} \sin \frac{\omega \ep_n}{2} \sin \frac{\omega\ep_{n-1}}{2} \ .
\ee

\begin{cor}
The equation
\be  \label{exact-drv1}
x_{n+1} - 2 x_n \cos\omega\ep + x_{n-1} = \frac{4 g}{\omega^2} \sin^2 \frac{\omega\ep}{2} 
\ee
is the exact discretization of \rf{osc-force} with a constant time step $\ep$. 
\end{cor}

\begin{prop}
The equation \rf{exact-drv1} can be rewritten in the following equivalent forms:
\be  \label{exact-oscdrv}
\frac{x_{n+1} - 2 x_n + x_{n-1}}{\left( \frac{2}{\omega} \sin \frac{\ep \omega}{2} \right)^2 } 
+ \omega^2 x_n  = g \ , 
\ee

\be  \label{oscdrv-dd1}
\frac{  v_n - v_{n-1} \cos\omega\ep }{ \frac{\sin\omega\ep}{\omega} } + \omega^2 x_{n-1} = \frac{g}{\cos^2 \frac{\omega\ep}{2} } \ ,
\ee

\be  \label{oscdrv-dd2}
\frac{ \left( v_n - \frac{g}{\omega} \tan\frac{\omega\ep}{2} \right)  - \left( v_{n-1} - \frac{g}{\omega} \tan\frac{\omega\ep}{2} \right) \cos\omega\ep }{ \frac{\sin\omega\ep}{\omega} } + \omega^2 x_{n-1} = g \ ,
\ee
where $v_n$ is defined by \rf{vn}.

\end{prop}

\subsection{Discrete analogues of the energy integral}

The energy integral for the equation \rf{osc-force} is given by
\be  \label{E-drv}
  E = \frac{1}{2} {\dot x}^2 + \frac{1}{2} \omega^2 x^2 - g x \ . 
\ee

\begin{prop} \label{prop-Ekn-drv}
Assuming that $x_n$ satisfies \rf{exact-drv1} we define: 
\be
E^{(0)}_n = x_{n+1}^2 - 2 \cos\omega\ep \ x_n x_{n+1} + x_n^2 - (x_n + x_{n+1})  \left( \frac{2 \sin\frac{\omega\ep}{2}}{\omega} \right)^2 g   , 
\ee

\be
E^{(1)}_n := \frac{1}{2} \left( \frac{x_{n+1} -  x_n}{ \frac{2}{\omega} \sin \frac{\ep \omega}{2} } \right)^2  
+ \frac{1}{2} \omega^2 x_n x_{n+1} - \frac{x_n + x_{n+1}}{2} \ g   \ ,
\ee

\be
E^{(2)}_n := \frac{1}{2} \left( \frac{x_{n+1} - x_n \cos\omega\ep}{\frac{\sin\omega\ep}{\omega}} \right)^2  + \frac{1}{2} \omega^2 x_n^2 
- \frac{x_n + x_{n+1}}{2 \cos^2 \frac{\omega\ep}{2} } \ g \ , 
\ee

\be
E^{(3)}_n := \frac{1}{2} \left( \frac{x_{n+1} - x_n \cos\omega\ep}{\frac{\sin\omega\ep}{\omega}} - \frac{g}{\omega} \ \tan \frac{\omega\ep}{2} \right)^2  + \frac{1}{2} \omega^2 x_n^2 
-  g x_n \ .
\ee

Then, 
$E^{(k)}_{n+1} = E^{(k)}_n$ for any $k = 0,1,2,3$ and any $n \in \Z$. Moreover,  

\be  \ba{l}  \dis  \label{ident2}
E^{(1)}_n = \frac{1}{2} \left( \frac{\omega}{2 \sin\frac{\omega\ep}{2}} \right)^2 \ E^{(0}_n \ , \\[3ex]  \dis
E^{(2)}_n = \left( 1 + \tan^2 \frac{\omega\ep}{2} \right) E^{(1)}_n = \frac{1}{2} \left( \frac{\omega}{ \sin{\omega\ep}} \right)^2 \ E^{(0}_n \ , \\[3ex] \dis 
E^{(3)}_n = E^{(2)}_n + \left( \frac{g}{\omega} \tan\frac{\omega\ep}{2} \right)^2 \ , 
\ea \ee
\end{prop}

\begin{Proof}
The identities \rf{ident2} can be easily verified. Therefore it is enough to show that $E^{(0)}_{n} = E^{(0)}_{n-1}$. We compute: 
\[
E^{(0)}_{n} - E^{(0)}_{n-1} = \left( x_{n+1} - x_{n-1} \right)\left( x_{n+1} + x_{n-1} - 2 \cos\omega\ep \ x_n - \frac{4 g}{\omega^2} \sin^2 \frac{\omega\ep}{2}  \right)  , 
\]
which vanishes provided that \rf{exact-drv1} holds. 
\end{Proof}

Therefore all four quantities defined in Proposition~\ref{prop-Ekn-drv} can be interpreted as  discrete analogues of the energy integral.  They correspond to different equivalent forms of the exact 
discrete harmonic oscillator equation with a constant driving force. In particular, 
 $E^{(1)}$ seems to be natural in the framework of Mickens' approach \cite{Mic}. However, the best discrete analogue of the energy integral is given by $E^{(3)}$.  Indeed, $E^{(3)}$ is of the form 
\rf{E-drv} evaluated at $x = x_n$ and $\dot x = v_n$, where 
\be  \label{vn-drv}
   v_n = \frac{x_{n+1} - x_n \cos\omega\ep}{\frac{\sin\omega\ep}{\omega}} - \frac{g}{\omega} \ \tan \frac{\omega\ep}{2} \ ,
\ee
compare \rf{vn-var-drv}.

\subsection{Damped harmonic oscillator}

The  damped harmonic oscillator with a constant driving force  
\be  \label{damp}
\ddot x = - \omega_0^2 x - 2 \gamma \dot x - g 
\ee
can be reduced to the harmonic oscillator without damping. 

\begin{prop}
The transformation
\be   \ba{l}
X = e^{\gamma t} \left( x - x_e \right) \ ,
\ea \ee
where $x_e = - \omega_0^{-2} g$, reduces \rf{damp} to the harmonic oscillator equation
\be
   \ddot X + \omega^2 X = 0 \ , \qquad  \omega^2 = \omega_0^2 - \gamma^2 \ .
\ee
\end{prop}

\begin{Proof} Straightforward computation. 
\end{Proof}

Denoting $p = m \dot x$ and $P = m \dot X$ we express $P$ in terms of $x, p$, namely:  
\be
P = e^{\gamma t} \left( p + m \gamma (x - x_e) \right) \ . 
\ee
Therefore, the exact discretization of the equation \rf{damp} in terms of variables $X, P$ is given by   
\be  \label{XnPn} 
\m X_{n+1} \\ P_{n+1} \ema = \mm \cos\omega\ep &  (m \omega)^{-1} \sin\omega\ep \\  - m \omega \sin\omega\ep & \cos\omega\ep \ema \m X_n \\ P_n \ema
\ee
(compare \rf{dis-osc-explicit}). Finally, substituting 
\be \ba{l}
  X_n = e^{\gamma t_n} \left( x_n - x_e \right) \ ,  \\[2ex]
  P_n = e^{\gamma t_n} \left( p_n + m \gamma (x_n - x_e) \right) 
\ea \ee
into \rf{XnPn} we obtain the final corollary.  

\begin{cor}
The exact discretization of the damped harmonic oscillator equation with the constant driving force \rf{damp} is given by
\be \ba{l}
x_{n+1} = x_e +  e^{-\gamma \ep}  \left(  (x_n - x_e) \left( \cos\omega\ep   + \frac{\gamma}{\omega} \sin\omega\ep \right)  + \frac{ \sin\omega\ep}{\omega} \ p_n \right) , \\[2ex]
p_{n+1} = e^{-\gamma \ep}  \left(  p_n \left( \cos\omega\ep  - \frac{\gamma}{\omega} \sin\omega\ep  \right) - \left(\omega + \frac{\gamma^2}{\omega} \right) (x_n - x_e) \sin\omega\ep 
\right)  .
\ea \ee

\end{cor}

\section{Exact discretization of the multidimensional harmonic oscillator equation}

In this section we consider the equation
\be  \label{vec-osc-gen}
 \frac{d^2   x }{d t^2}  +  \Omega^2  x = f (t) \ , 
\ee
where $x = x(t) \in \R^n$, $ \Omega^2$ is a given invertible constant $n \times n$ matrix and $f = f(t) \in \R^n$ is a given driving force. This is a natural $n$-dimensional extension of the harmonic oscillator equation. Indeed, if $f = 0$ and $\Omega^2$ has $n$ pairwise different eigenvalues, then the equation \rf{vec-osc} describes $n$ independent one-dimensional harmonic oscillators. 

Here, in contrast to previous sections, we try to use the variable time-step whenever possible. 

\subsection{Multidimensional harmonic oscillator}

We begin with the case $f (t) \equiv 0$, i.e., 
\be  \label{vec-osc}
 \frac{d^2   x }{d t^2}  +  \Omega^2  x = 0 \ . 
\ee
It is convenient to represent 
\rf{vec-osc} as the following first order system
\be  \label{vec1-osc}
  \dot x = v \ , \quad \dot v = - \Omega^2 x \ . 
\ee
 The general exact solution is given by
\be \ba{l}  \label{vo-sol}
 x (t)  = e^{i \Omega t}  c_1 + e^{- i \Omega t}  c_2 \ , \\[2ex]
v (t) = i \Omega \left( e^{i \Omega t}  c_1 - e^{- i \Omega t}  c_2 \right) \ , 
\ea \ee
where $ c_1$, $ c_2$ are constant $n$-vectors. 

The exact discretization of the equation \rf{vec-osc} is given by
$ x_n :=  x (t_n)$, where we use \rf{vo-sol}. Therefore
\be \ba{l}  \label{von}
 x_n  = e^{i t_n \Omega }  c_1 + e^{- i t_n \Omega t}  c_2 \ , \\[2ex]
 v_n = i \Omega \left( e^{i t_n \Omega}  c_1 - e^{- i t_n \Omega }  c_2 \right) \ .
\ea \ee
Hence
\be  \label{von'}
e^{i t_n \Omega }  c_1 = \frac{1}{2} \left( x_n - i \Omega^{-1} v_n \right) \ , \quad 
e^{- i t_n \Omega }  c_2 = \frac{1}{2} \left( x_n + i \Omega^{-1} v_n \right) \ , \quad 
\ee
and, denoting $\ep_n = t_{n+1} - t_n$ (the time step, in general variable), we have
\be \ba{l}  \label{von1}
 x_{n+1}  = e^{i t_n \Omega + i \ep_n \Omega }  c_1 + e^{- i t_n \Omega t - i \ep_n \Omega }  c_2 \ , \\[2ex]
 v_{n+1} = i \Omega \left( e^{i t_n \Omega + i \ep_n \Omega }  c_1 - e^{- i t_n \Omega - i \ep_n \Omega  }  c_2 \right) \ .  
\ea \ee
Substituting \rf{von'} into \rf{von1} we obtain:
\be  \label{osc-exact}
\left( \ba{c}   x_{n+1} \\  v_{n+1} \ea \right) = \left( 
\ba{cc}   \cos  \Omega \ep_n & { \Omega}^{-1} \sin  \Omega \ep_n \\ 
-  \Omega \sin  \Omega \ep_n  & \cos  \Omega \ep_n  \ea \right) 
\left( \ba{c}   x_n \\  v_n \ea \right) \ ,
\ee
where we use a natural matrix notation (the matrix entries and vector components  are $n\times n$ matrices). In particular, we have:
\be \label{vnep} 
 v_n = \Omega (\sin\Omega \ep_n)^{-1} \left( x_{n+1} - (\cos\Omega \ep_n) x_n \right) \ . 
\ee
The matrix on the right-hand side of \rf{osc-exact} is an even function of $\Omega$, i.e., this matrix can be expressed as a series in terms of $\Omega^2$ (without explicit knowledge of $\Omega$).

\begin{cor}
The exact discretization of the system \rf{vec1-osc} is given by \rf{osc-exact}, where $\ep_n$ is an arbitrary variable time step (in particular, we can take $\ep_n = \ep = \const$).
\end{cor}

Below we present three propositions. The proofs are omitted because in Section~\ref{sec-multi-drv} we will prove more general theorems.

\begin{prop}  \label{prop-multi-osc}
The exact discretization \rf{osc-exact} of the multi-dimensional harmonic oscillator equation \rf{vec-osc} is equivalent to the system
\be \ba{l} \label{multi-osc}
\delta_n^{-1} \left(   x_{n+1} -   x_n \right) = \frac{1}{2} \left(  v_{n+1} + 
 v_n \right) \ , \\[2ex]
\delta_n^{-1} \left(  v_{n+1} -  v_n \right) = - \frac{1}{2} \Omega^2 \left(   x_{n+1} + 
  x_n \right) 
\ea \ee
where 
\be  \label{deltaOM} 
 \delta_n = 2 \Omega^{-1} \tan \frac{\Omega \ep_n}{2} \ . 
\ee
\end{prop}

\begin{prop}
If the time step is constant ($\ep_n = \ep$), then the exact discretization \rf{osc-exact} of the multi-dimensional harmonic oscillator equation \rf{vec-osc} is equivalent to the system
\be \ba{l}   \label{vo-sec}
 x_{n+1} - 2 (\cos \Omega \ep ) \ x_n + x_{n-1} = 0 \ , \\[2ex] 
v_n = \Omega (\sin\Omega \ep)^{-1} \left( x_{n+1} - (\cos\Omega \ep) x_n \right) \ .
\ea \ee
\end{prop}

\begin{prop} 
If  \ $v_n$ is defined by \rf{vnep} and \ $\Omega^T = \Omega$, then  
\be
I_n : = \frac{1}{2} { |v_n|}^2 + \frac{1}{2} \scal{ x_n}{ \Omega^2  x_n} 
\ee 
is an integral of motion of the discrete multidimensional 
harmonic oscillator equations \rf{multi-osc} (i.e., \ $I_{n+1} = I_n$). 
\end{prop}

Here (and below)  $\scal{\phi}{\psi}$ denotes a scalar product in $\R^n$, $|\phi|^2 = \scal{\phi}{\phi}$ and  $\Omega^T$ is the transpose of $\Omega$. We recall that 
\be  \label{skaltran}
   \scal{ M \phi}{\psi} = \scal{\phi}{M^T \psi}  \quad {\rm and} \quad  \scal{\phi}{\psi} = \scal{\psi}{\phi} \ 
\ee
for any matrix $M$ and any vectors $\phi, \psi \in \R^n$. Moreover, it is worthwhile to point out the obvious fact that $\Omega$, $\delta_n$, $\sin\Omega\ep_n$ (and other analytic functions of $\Omega$) pairwise commute. We use frequently this property.

\subsection{Multidimensional harmonic oscillator with a constant driving force}
\label{sec-multi-drv}

We consider the equation
\be  \label{vec-osc2}
 \frac{d^2   x }{d t^2}  +  \Omega^2  x = a \ , 
\ee
where $x = x(t) \in \R^n$ and $ \Omega$ is a given invertible constant $n \times n$ matrix and $a = \const \in \R^n$. It is convenient to represent 
\rf{vec-osc2} as the following first order system
\be  \label{vec1-osc2}
  \dot x = v \ , \quad \dot v = - \Omega^2 x + a \ . 
\ee
 Its general exact solution is given by
\be \ba{l}  \label{vo-sol2}
 x (t)  = e^{i \Omega t}  c_1 + e^{- i \Omega t}  c_2 + \Omega^{-2} a \ , \\[2ex]
v (t) = i \Omega \left( e^{i \Omega t}  c_1 - e^{- i \Omega t}  c_2 \right) \ , 
\ea \ee
where $ c_1$, $ c_2$ are constant $n$-vectors. 

The exact discretization of the equation \rf{vec-osc2} is given by
$ x_n :=  x (t_n)$, where we use \rf{vo-sol2}. Therefore
\be \ba{l}  \label{von2}
 x_n  = e^{i t_n \Omega }  c_1 + e^{- i t_n \Omega t}  c_2 + \Omega^{-2} a \ , \\[2ex]
 v_n = i \Omega \left( e^{i t_n \Omega}  c_1 - e^{- i t_n \Omega }  c_2 \right) \ .
\ea \ee
Hence
\be \ba{l} \label{von''} \dis
e^{i t_n \Omega }  c_1 = \frac{1}{2} \left( x_n - \Omega^{-2} a  - i \Omega^{-1} v_n \right) \ , \\[3ex] \dis 
e^{- i t_n \Omega }  c_2 = \frac{1}{2} \left( x_n - \Omega^{-2} a + i \Omega^{-1} v_n \right) \ .  
\ea \ee
Denoting $\ep_n = t_{n+1} - t_n$ (the time step, in general variable) and evaluating \rf{von2} at $n+1$, we obtain 
\be \ba{l}  \label{von3}
 x_{n+1}  = e^{i t_n \Omega + i \ep_n \Omega }  c_1 + e^{- i t_n \Omega t - i \ep_n \Omega }  c_2 + \Omega^{-2} a \ , \\[2ex]
 v_{n+1} = i \Omega \left( e^{i t_n \Omega + i \ep_n \Omega }  c_1 - e^{- i t_n \Omega - i \ep_n \Omega  }  c_2 \right) \ .  
\ea \ee

\ods

\begin{prop}
The exact discretization of the system \rf{vec1-osc2} is given by 
\be  \label{osc-exact2}
\left( \ba{c}   x_{n+1} \\  v_{n+1} \ea \right) = \left( 
\ba{cc}   \cos  \Omega \ep_n & { \Omega}^{-1} \sin  \Omega \ep_n \\ 
-  \Omega \sin  \Omega \ep_n  & \cos  \Omega \ep_n  \ea \right) 
\left( \ba{c}   x_n \\  v_n \ea \right) + \m 2 \Omega^{-2} \sin^2\frac{\Omega\ep_n}{2} \ a \\ \Omega^{-1} \sin\Omega\ep_n \ a \ema   , 
\ee
where $\ep_n$ is an arbitrary variable time step (in particular, we can take $\ep_n = \ep = \const$).
\end{prop}

\begin{Proof} It is enough to 
substitute \rf{von''} into \rf{von3}. 
\end{Proof}
\ods

In particular,
\be  \label{vnepa}
v_n = \Omega (\sin\Omega \ep_n )^{-1} \left( x_{n+1} - (\cos\Omega \ep_n ) x_n \right) - \Omega^{-1} \tan \frac{\Omega\ep_n }{2} \ a \ .
\ee

\ods

\begin{prop}  \label{prop-a-good}
The exact discretization \rf{osc-exact2} of the multi-dimensional harmonic oscillator equation with the constant driving force \rf{vec1-osc2} is equivalent to the system
\be \ba{l} \label{multi-osc2}
\delta_n^{-1} \left(   x_{n+1} -   x_n \right) = \frac{1}{2} \left(  v_{n+1} + 
 v_n \right) \ , \\[2ex]
\delta_n^{-1} \left(  v_{n+1} -  v_n \right) = - \frac{1}{2} \Omega^2 \left(   x_{n+1} + 
  x_n \right) + a \ ,
\ea \ee
where $\delta_n$ is given by \rf{deltaOM}. 
\end{prop}

\begin{Proof}
From \rf{osc-exact2} we compute:
\be  \ba{l}
 x_{n+1} + x_n =  (1 + \cos\Omega\ep_n) x_n + (\Omega^{-1} \sin\Omega\ep_n) v_n  + (1 - \cos\Omega\ep_n) \Omega^{-2} a \ , \\[3ex] 
 x_{n+1} - x_n =  - (1 - \cos\Omega\ep_n) x_n + (\Omega^{-1} \sin\Omega\ep_n) v_n  + (1 - \cos\Omega\ep_n) \Omega^{-2} a \ , \\[3ex]
v_{n+1} + v_n = - (\Omega \sin\Omega\ep_n) x_n +  (1 + \cos\Omega\ep_n) v_n + \Omega^{-1} \sin\Omega\ep_n \ a \ , \\[3ex] 
v_{n+1} - v_n = - (\Omega \sin\Omega\ep_n) x_n - (1 - \cos\Omega\ep_n) v_n + \Omega^{-1} \sin\Omega\ep_n \ a \ . 
\ea \ee
Hence
\be \ba{l}  \dis
 \Omega \sin\frac{\Omega\ep_n}{2} \left( x_{n+1} + x_n \right) +  \cos\frac{\Omega\ep_n}{2} \left(  v_{n+1} - v_n \right) =  2 \Omega^{-1} \sin \frac{\Omega\ep_n}{2} \ a \ , \\[3ex]
\Omega \cos\frac{\Omega\ep_n}{2} \left( x_{n+1} - x_n \right) =   \sin\frac{\Omega\ep_n}{2} \left(  v_{n+1} + v_n  \right) \ , 
\ea \ee
which is equivalent to \rf{multi-osc2}. 
\end{Proof}

\begin{prop} 
If the time step is constant ($\ep_n = \ep$), then the exact discretization \rf{osc-exact2} of the multidimensional harmonic oscillator with a constant  force can be rewritten in the following equivalent form
\be \ba{l}   \label{vo-sec2} \dis
 x_{n+1} - 2 (\cos \Omega \ep ) \ x_n + x_{n-1} = \left(  2  \Omega^{-1} \sin\frac{\Omega\ep }{2} \right)^2 a \ , \\[3ex]\dis
v_n = \Omega (\sin\Omega \ep )^{-1} \left( x_{n+1} - (\cos\Omega \ep ) x_n \right) - \Omega^{-1} \tan \frac{\Omega\ep }{2} \ a \ .
\ea \ee
Another convenient expression for the velocity $v_n$ is given by:
\be  \label{vsred}
v_n = \frac{1}{2} \Omega (\sin\Omega\ep)^{-1} \left( x_{n+1} - x_{n-1} \right) \ .  
\ee
\end{prop}

\begin{Proof} From \rf{osc-exact2} we get:
\be \ba{l}  \label{vsys}
\Omega^{-1} (\sin\Omega\ep) v_n = \left( x_{n+1} - (\cos\Omega \ep ) x_n \right) - \Omega^{-1} \tan \frac{\Omega\ep }{2} \ a \ , 
\\[2ex] 
v_n = - \Omega \sin\Omega\ep \ x_{n-1} + \cos\Omega\ep \ v_{n-1} + \Omega^{-1} \sin\Omega\ep \ a \ ,  \\[2ex] 
v_{n-1} = \Omega (\sin\Omega \ep )^{-1} \left( x_n - (\cos\Omega \ep ) x_{n-1} \right) - \Omega^{-1} \tan \frac{\Omega\ep }{2} \ a \ .
\ea \ee 
Thus we derived the formula \rf{vo-sec2} for $v_n$.  
Eliminating $v_{n}$ and $v_{n-1}$ from the system \rf{vsys} we get the first equation of \rf{vo-sec2}. In order to obtain the formula \rf{vsred} we evaluate \rf{von2} at $n\pm 1$
\be
x_{n\pm 1} = e^{i t_n \Omega \pm i \ep_n \Omega }  c_1 + e^{- i t_n \Omega t \mp i \ep_n \Omega }  c_2 + \Omega^{-2} a \ , 
\ee
and compute
\be
\frac{1}{2} \left( x_{n+1} - x_{n-1} \right)  = i e^{t_n \Omega} (\sin\ep_n\Omega) c_1 - i e^{-i t_n \Omega} (\sin\ep_n\Omega) c_2 \ .
\ee
Taking into account the second equation of \rf{von2} we complete the proof.  \end{Proof}

\ods

\begin{prop}  \label{prop-energy-multi-a}
If \ $\Omega^T = \Omega$ and $v_n$ is defined by \rf{vnepa}, then 
\be  \label{Ina}
I_n : = \frac{1}{2} { |v_n| }^2 + \frac{1}{2} \scal{ x_n}{ \Omega^2  x_n} - \scal{a}{ x_n } 
\ee 
is an integral of motion (i.e., $I_{n+1} = I_n$) of the discrete multidimensional 
harmonic oscillator equations \rf{multi-osc2}. 
\end{prop}

\begin{Proof}
From \rf{multi-osc2} we have
\be \ba{l} \label{multi-osc2'}
  v_{n+1} + 
 v_n = 2 \delta_n^{-1} \left(   x_{n+1} -   x_n \right) \ , \\[2ex]
 v_{n+1} -  v_n  = - \frac{1}{2} \delta_n \Omega^2 \left(   x_{n+1} + 
  x_n \right) + \delta_n a \ . 
\ea \ee 
In order to prove $I_{n+1}=I_n$ it is enough to multiply the above  equations side by side (using the scalar product and its properties). $\Omega^T = \Omega$ obviously implies $\delta_n^T = \delta_n$. Taking also into account \rf{skaltran}, we verify that 
\be
\scal{\delta_n^{-1} x_{n+1}}{\delta_n\Omega^2 x_n} = 
\scal{\delta_n^{-1} x_n}{\delta_n\Omega^2 x_{n+1}} \ .
\ee
Therefore, multiplying \rf{multi-osc2'} side by side, we get
\be
 |v_{n+1}|^2 -  |v_n|^2 = \scal{\delta_n^{-1} x_{n+1}}{\delta_n \Omega^2 x_{n+1}} - \scal{\delta_n^{-1} x_n}{\delta_n\Omega^2 x_n} + 2 \scal{x_{n+1} - x_n}{a} \ . 
\ee
Hence, applying once more \rf{skaltran}, we get $I_{n+1} = I_n$. 
\end{Proof}

\subsection{Multidimensional harmonic oscillator with a polynomial driving force}

In this section we consider the harmonic oscillator perturbed by an arbitrary polynomial driving force:
\be  \label{poldrive} 
\ddot x + \Omega^2 x = f (t) \ , \quad f (t) = \sum_{k=0}^N c_k t^k \ ,
\ee
where $x = x (t) \in \R^n$, $ f (t) \in \R^n$, $c_N \neq 0$ and $\Omega^2$ is a constant invertible matrix.  

The key observation is that the general solution to \rf{poldrive} can be represented in the form
\be  \label{xfi}
  x (t) = (\cos\Omega t) A + (\sin\Omega t) B + \Phi (t) \ , 
\ee
where where $A, B \in \R^n$ and 
\be  \label{fi} 
\Phi (t) := \Omega^{-2} \sum_{k=0}^\infty (-\Omega^{-2})^k f^{(2 k)} (t) \ .
\ee
One can easily see that the infinite sum contains only a finite number, $[N/2]$, of non-zero terms and  $\Phi (t)$ is a polynomial  of $N$th order. Moreover, 
\be  \label{vfi}
 v (t) \equiv \dot x (t) = \Omega (\cos\Omega t) B - \Omega (\sin\Omega t) A + \dot \Phi (t) \ .
\ee

\begin{prop}
The exact discretization of \rf{poldrive} is given by
\be  \ba{l} \label{exact-polydrive} 
x_{n+1} - 2 \cos\Omega\ep \ x_n + x_{n-1} = \Phi_{n+1} - 2 (\cos\Omega\ep ) \Phi_n  + \Phi_{n-1} \ , \\[2ex]
v_n = \frac{1}{2} \Omega (\sin\Omega\ep)^{-1} \left( x_{n+1} - x_{n-1} - \Phi_{n+1} + \Phi_{n-1}  \right) + {\dot\Phi}_n \ , 
\ea \ee
where $\Phi_n = \Phi (n\ep)$, $\dot\Phi_n = \dot\Phi (n \ep)$ and $\Phi (t)$ is defined by \rf{fi}. 
\end{prop}

\begin{Proof}
Computing $x (t\pm \ep)$ and expressing $A, B$ in terms of $x, v, \Phi, \dot \Phi$  we get: 
\be \ba{l}
  x (t\pm \ep) = (\cos\Omega\ep) ( (x (t) - \Phi(t)) \pm \Omega^{-1} (\sin\Omega\ep) ( v (t) - \dot \Phi (t)) + \Phi (x\pm\ep)  ,
\\[2ex]
 x (t+\ep) + x (t-\ep) = 2 (\cos\Omega\ep ) ( x (t) - \Phi (t)) + \Phi (t + \ep) + \Phi (t-\ep) \ , \\[2ex]
x (t+\ep) - x (t-\ep) = 2 \Omega^{-1} (\sin\Omega\ep) ( v (t) - \dot\Phi (t) ) + \Phi (t+\ep) - \Phi (t-\ep) \ .
\ea \ee
Now, identifying $t \rightarrow n\ep$, $x (n\ep) \rightarrow x_n$, $\Phi (n\ep) \rightarrow \Phi_n$, we obtain the exact discretization of the perturbed harmonic oscillator equation \rf{exact-polydrive}. 
\end{Proof}

In the particular case $f (t) \equiv a = \const$, we have
\be \ba{l}
\Phi (t) = \Omega^{-2} f = \Omega^{-2} a \ , \\[2ex] 
\Phi_{n+1} - 2 \cos\Omega\ep \Phi_n + \Phi_{n-1} = 2 (1 - \cos\Omega\ep) \Omega^{-2} a \ . 
\ea \ee
Therefore, in this case \rf{exact-polydrive} reduces to \rf{vo-sec2}. 

\ods

Analogical considerations can be made for any other perturbation  (driving force) $f = f(t)$ such that the series 
\be
  \sum_{k=0}^{\infty} (-\Omega^{-2})^k f^{2 k} (t) 
\ee
is finite or summable. For instance, in some cases we  get geometric series. If $f (t) = e^{\alpha t} f_0 $, ($\alpha = \const \in \R$), then 
\be
\Phi (t) = \Omega^{-2} \sum_{k=0}^\infty (- \alpha^2 \Omega^{-2})^k e^{\alpha t} f_0  =  \left( \Omega^2 + \alpha^2 {\mathbb I} \right)^{-1} e^{\alpha t} f_0 \ ,
\ee
where $\mathbb I$ is the unit matrix in $\R^n$. 
If $f (t) = f_0 \sin\omega t$, ($\omega = \const \in \R$), then 
\be
\Phi (t) = \Omega^{-2} \sum_{k=0}^\infty ( \omega^2 \Omega^{-2})^k (\sin\omega t) f_0  =  \left( \Omega^2 - \omega^2 {\mathbb I} \right)^{-1} e^{\alpha t} f_0 \ . 
\ee
Using the linearity of \rf{fi} one can extend these results on 
finite Fourier series,  finite linear combination of exponential functions or some specific combinations  of polynomials, exponentials and triginometric functions.  Similar  results were first obtained (in a different way) a long time ago \cite{Be1,Be2}.

\section{Connections with numerical methods} 
\label{sec-numer} 

In order to produce an exact discretization we have to know corresponding  exact solutions and in such cases numerical schemes seem to be of a little practical use. It turns out,  however, that exact discretizations can be applied to construct  good numerical schemes which are exact in some particular cases.

\subsection{Gautschi-type methods} 

The exact discretization of the harmonic oscillator is of great importance for a large class of numerical methods (Gautschi-type methods) \cite{Gau,HoL2}. In order to illustrate this idea we 
consider the following class of equations: 
\be
\ddot x = - \omega^2 x + g (x) \ , 
\ee
where $x \in \R^n$, $\omega \in \R$ and the nonlinear term $ g (x)$ is assumed to be much smaller than the linear term $\omega^2 x$.  The simplest Gautschi method is defined as \cite{Gau,HLW}:
\be
x_{n+1}  - 2 (\cos \omega\ep) x_n + x_{n-1} = \left( \frac{2}{\omega} \sin \frac{\omega\ep}{2} \right)^2 g (x_n)  \ .
\ee
One can easily see that for $g (x) = \const$ the Gautschi method \cite{Gau} reduces to the exact discretization of the harmonic oscillator with a constant driving force, compare \rf{exact-drv1}. In other cases the Gautschi method ceased to be exact but is still considered as one of the best for oscillatory problems with constant high frequencies \cite{HLW,HoL2}. 

\subsection{Exponential integrators}

Exponential integrators \cite{MW,Pope} are also concerned with exact discretizations. We confine ourselves to equations of the form
\be  \label{Lg} 
   \dot y = L y + g (y) \ , 
\ee
where $y \in \R^n$, $L$ is a constant matrix and $g (y)$ is a small nonlinear term. We consider three exponential integrators which are extensions of the classical Euler method: the explicit Euler-Lawson scheme   
\be  \label{eEL}
y_{n+1} = e^{\ep L} \left(  y_n + \ep g (y_n) \right)  \  , 
\ee
the implicit Euler-Lawson scheme   
\be
y_{n+1} = e^{\ep L}  y_n + \ep g (y_{n+1})   \ , 
\ee
and the exponential Euler method:
\be  \label{exE}
y_{n+1} = e^{\ep L} y_n +  \ep \ph_1 (\ep L) g (y_n) \  , 
\ee
where $\ph_1 (z) := (e^z - 1)/z$. The equation \rf{osc-force} obviously can be represented in the form \rf{Lg}:
\be  \label{Lg-osc}
\frac{d}{d t} \m x \\ v \ema = \mm 0 & 1 \\ -\omega^2 & 0 \ema \m x \\ v \ema + \m 0 \\ g \ema  \ ,
\ee
where $g = \const$. 
Applying the above exponential integrators to \rf{Lg-osc} we get, respectively, 
\be  \ba{l}  
\m x_{n+1} \\ v_{n+1} \ema = \mm \cos\omega\ep & \omega^{-1} \sin\omega\ep \\ - \omega\sin\omega\ep & \cos\omega\ep \ema   \m x_n \\ v_n \ema  + \frac{\ep g}{\omega} \m  \sin\omega\ep \\ \omega\cos\omega\ep  \ema  , 
\\[4ex] 
\m x_{n+1} \\ v_{n+1} \ema = \mm \cos\omega\ep & \omega^{-1} \sin\omega\ep \\ - \omega\sin\omega\ep & \cos\omega\ep \ema   \m x_n \\ v_n \ema  + \m 0 \\ \ep g \ema ,
\\[4ex] 
\m x_{n+1} - \omega^{-2} g  \\ v_{n+1} \ema = \mm \cos\omega\ep & \omega^{-1} \sin\omega\ep \\ - \omega\sin\omega\ep & \cos\omega\ep \ema   \m x_n  - \omega^{-2} g \\ v_n \ema  . 
\ea \ee
We see that in the case $g = 0$ (the harmonic oscillator without the driving force) all three integrators yield the exact discretization \rf{dis-osc-explicit}. In the case $g \neq 0$ only the exponential Euler scheme \rf{exE} is exact, compare \rf{dis-osc-explicit-g}. 

\subsection{Application to the Kepler motion} 

Sometimes considered equations reduce to the harmonic oscillator (or contain the harmonic oscillator equation).  Then we may take advantage of using the exact discretization of the harmonic oscillator. Here we shortly describe the case of the classical Kepler problem 
\be
 m \ddot {\mathbf r} = - \frac{k {\mathbf r}}{r^3} \ , \qquad {\mathbf r} \in \R^3 \ , \qquad r = |\mathbf r| \ , 
\ee
where $m, k$ are  constant. 
It is well known (see, for instance, \cite{HLW,RK}), that exact trajectories are given by $r = r (\ph)$, where $\ph$ is the angle swept out by $\mathbf r$ and 
$1/r$ solves the so called Binet equation (equivalent to the 
harmonic oscillator equation with the constant driving force): 
\be \label{Bin} 
      \frac{d^2 u}{d \ph^2} + u  =   \frac{k m}{L^2} \ , \qquad u = \frac{1}{r} \ , 
\ee 
where $L = |{\mathbf r} \times m \dot{\mathbf r}| = \const$ (the angular momentum) is an integral of motion. Using 
the exact discretization of the harmonic oscillator equation \rf{Bin} we get a numerical scheme such that $\ph_{n+1} - \ph_n = \const$ (but the time step is variable) \cite{Ci-Kep}.  

It is also well known \cite{KS} that the Kepler problem can be reduced to the 4-dimensional harmonic oscillator equation 
\be  \label{Q-KS} 
 \frac{d^2 Q}{d s^2} - \frac{E}{2} Q = 0 \ , \qquad Q = Q (s) \in \R^4 \ , 
\ee
where $E$ is the energy integral of the considered Kepler motion.   Having $Q$ and $s$ we can find $\mathbf r$ (explicitly expressed in terms of $Q$) and $t$ (integrating the equation $t' = |Q|^2$), for details and precise formulas see, for instance, \cite{Koz,KS,MN}. 

However, the possibility of using the exact discretization has been often overlooked. First, Minesaki and Nakamura \cite{MN} discretized the harmonic oscillator  equation \rf{Q-KS} by the discrete gradient method \cite{Gree,LaG}. The obtained discretization was quite good (the exact trajectories were preserved) but not exact. Then, Kozlov \cite{Koz} succeeded to get the exact discretization by summing up some infinite series. 
Earlier papers \cite{Be2,SB}, where the exact discretization 
of the harmonic oscillator was taken into account, now seem to be  forgotten (as far as geometric numerical integration is concerned). More detailed discussion of the exact discretization of the Kepler motion based on the Kustaanheimo-Stiefel theory can be found in \cite{Ci-Koz}.

\subsection{Locally exact modification of the discrete gradient  scheme} 
\label{sec-loc-exact}

Locally exact numerical schemes generalize the well known concept of the small oscillations approximation \cite{Ci-focm,CR-long,CR-grad}. If a numerical method contains some parameters, we modify these parameters so that the resulting scheme is exact for small oscillations \cite{CR-long} or, in more general setting, we require that the resulting scheme is exact for the linearization of the considered system \cite{Ci-focm,CR-grad}. 
As an illustrative example we consider the system
\be \label{Newton}
  \dot v = - \Phi' (x) \ ,  \quad v = \dot x \ ,
\ee
where $\Phi = \Phi (x) \in \R$ is a given function. The case $\Phi (x) = \frac{1}{2} \omega^2 x^2 + g x$ corresponds to the harmonic oscillator with a constant driving force. The energy conservation law reads
\be
  \frac{1}{2} v^2 + \Phi (x) = \const \ .
\ee 
The discrete gradient method \cite{Gree,LaG,MQR2} for \rf{Newton} yields: 
\be \ba{l} \label{dis-grad} \dis 
\frac{v_{n+1} - v_n}{\ep} =  - \frac{ \Phi (x_{n+1}) - 
\Phi (x_n) }{x_{n+1} - x_n} \ , \\[3ex] \dis
 \frac{1}{2} \left( v_{n+1} + v_n \right) = \frac{x_{n+1} - x_n}{\ep} 
 \ .
\ea \ee
We consider the following extension of the discrete gradient scheme: 
\be \ba{l} \label{mod-dis-grad} \dis
\frac{v_{n+1} - v_n}{\delta_n} =  - \frac{ \Phi (x_{n+1}) - 
\Phi (x_n) }{x_{n+1} - x_n} \ , \\[3ex] \dis
 \frac{1}{2} \left( v_{n+1} + v_n \right) = \frac{x_{n+1} - x_n}{\delta_n} 
 \ , 
\ea \ee
where $\delta_n$ is an arbitrary positive function of $\ep_n, x_n, v_n, x_{n+1}, v_{n+1}$ etc. The system \rf{mod-dis-grad} is a consistent approximation of \rf{Newton} if we add the natural consistency condition
\be  \label{consis} 
  \lim_{\ep \rightarrow 0} \frac{\delta_n}{\ep} = 1  \ . 
\ee 
\ods

\begin{prop} 
The numerical scheme \rf{mod-dis-grad} preserves exactly the energy integral (for any positive function $\delta_n$). 
\end{prop}

\begin{Proof}
We multiply side by side both equations of \rf{mod-dis-grad} obtaining:
\be
\frac{1}{2} v_{n+1}^2 + \Phi (x_{n+1}) = \frac{1}{2} v_n^2 + \Phi (x_n) \ ,
\ee
which ends the proof. 
\end{Proof}

In order to find the most accurate discretizations among \rf{mod-dis-grad} we linearize \rf{mod-dis-grad} around $x = \bar x$ (and $\bar x$ will be specified below):
\be \ba{l} \label{dis-grad-approx} \dis
\frac{\xi_{n+1} - \xi_n}{\delta_n} = \frac{1}{2} \left( p_{n+1} + p_n \right) \ . \\[3ex] \dis
\frac{p_{n+1} - p_n}{\delta_n} =  - \Phi'(\bar x) - \frac{1}{2} \Phi'' (\bar x) \left( \xi_n + \xi_{n+1} \right) 
\ , 
\ea \ee
where $\xi_n := x_n - \bar x$ and $\xi_{n+1} = x_{n+1} - \bar x$. 
The resulting system \rf{dis-grad-approx} is equivalent to the harmonic oscillator equation with a constant driving force, compare  
\rf{osc-graddel}, provided that we identify $p_n = v_n$ and require
\be
  g = - \Phi' (\bar x) \ , \qquad \omega^2 = V'' (\bar x) 
\ , \qquad    \delta_n = \frac{2}{\omega} \tan \frac{\omega\ep}{2} \ 
\ee 
(note that the consistency condition \rf{consis} is obviously satisfied). 
Choosing $\bar x = x_n$ or $\bar x = \frac{1}{2} (x_n + x_{n+1})$ (thus changing $\bar x$ at every step) we obtain two very accurate (``locally exact'') numerical schemes, compare \cite{CR-grad}. 

\subsection{The wave equation}

In this section we are going to show that the exact discretization of the harmonic oscillator equation can be applied also in the case of partial differential equations.  We consider the linearized wave equation
\be
u_{tt} = u,_{xx} - a^2 u \ . 
\ee
Folowing Bridges and Reich (\cite{BR}, Section~4.2) we perform the Fourier transformation, substituting  $u (x, t) = {\hat u} (t) e^{ikx}$. Hence  
\be  
  \frac{d^2}{d t^2} \hat u = - \omega^2 \hat u \ , 
\ee 
where $\omega^2 = k^2 + a^2$. Instead of using standard methods (like implicit midpoint/trapezoidal schemes, see \cite{BR}) we take advantage of using the exact discretization:
\be
 {\hat u}^{n+1} - 2 (\cos\omega \Delta t) {\hat u}^n + {\hat u}^{n-1} = 0 \ . 
\ee
Thus the ``numerical frequency'' of the solution ${\hat u}^n$ (see \cite{BR}) coincides with $\omega$. It leads to many important improvements. For instance, the group velocity becomes exact and our numerical scheme reproduces exactly the energy transport in wave packets. 

\section{Conclusions and future directions}
\label{sec-wave}

The exact discretization of linear ordinary differential equations is a subject rather well known but a little bit forgotten. In this paper we have shown a large number of useful properties and equivalent formulations of the exact discretizations of harmonic oscillator equation and its variuos extensions and generalizations. Among these results it is worthwhile to distinguish Propositions~\ref{prop-multi-osc} and \ref{prop-a-good}, with formulas \rf{multi-osc} and \rf{multi-osc2}, and Proposition~\ref{prop-energy-multi-a} (the energy conservation law in a quite general case). 

Our main motivation for careful studies of exact discretizations consists in potential numerical applications. This point is explained in Section~\ref{sec-numer}. In particular, a new concept for modifying numerical schemes is presented in Section~\ref{sec-loc-exact}.  
We point out that this 
derivation of the locally exact discrete gradient method  is short and much simpler than the presentation given in our paper \cite{CR-grad}. This simplification is due to the direct use of the formula \rf{osc-graddel} which is especially suitable for any applications related to the discrete gradient method. 
Other results of this paper turn out to be very helpful in deriving and studying other locally exact numerical schemes, to be published soon \cite{Ci-focm}

A promising future direction is associated with the time scale approach \cite{BP,Hi}. Its main goal is to unify differential and difference calculus. Exact discretizations also connect smooth and discrete case but in an apparently different way. We plan to find links between these two approaches. A natural possibility is to generalize the notion of the delta derivative using new difference operators, e.g., the difference operator $\Delta_\ep$, see \rf{newDelta}. 
Exact discretizations of the harmonic oscillator equation can be also applied in the case of operators on Banach spaces (see \cite{CCL}) and for some PDEs (see Section~\ref{sec-wave}). In the near future we plan to study in more detail the exact discretization of the wave equation.


\end{document}